\documentclass[conference]{IEEEtran}
\IEEEoverridecommandlockouts
\usepackage{cite}
\usepackage{amsmath,amssymb,amsfonts}
\usepackage{algorithmic}
\usepackage{graphicx}
\usepackage{textcomp}
\usepackage{xcolor}
\usepackage{lipsum}
\usepackage{subcaption}
\usepackage{hyperref}
\usepackage{multirow}
\def\BibTeX{{\rm B\kern-.05em{\sc i\kern-.025em b}\kern-.08em
    T\kern-.1667em\lower.7ex\hbox{E}\kern-.125emX}}
\begin{document}

\title{Identifying and Calibrating Overconfidence in Noisy Speech Recognition \\
}


\author{
\IEEEauthorblockN{
Mingyue Huo\IEEEauthorrefmark{1},
Yuheng Zhang\IEEEauthorrefmark{2},
Yan Tang\IEEEauthorrefmark{1}\IEEEauthorrefmark{3}
}
\IEEEauthorblockA{\IEEEauthorrefmark{1}Department of Linguistics,
\IEEEauthorrefmark{2}Department of Computer Science, 
University of Illinois Urbana-Champaign, Urbana, USA}
\IEEEauthorblockA{\IEEEauthorrefmark{3}Beckman Institute for Advanced Science and Technology, University of Illinois Urbana-Champaign, Urbana, USA}
\IEEEauthorblockA{\{mhuo5, yuhengz2, yty\}@illinois.edu}
}


\maketitle

\begin{abstract}
Modern end-to-end automatic speech recognition (ASR) models like Whisper not only suffer from reduced recognition accuracy in noise, but also exhibit overconfidence---assigning high confidence to wrong predictions. We conduct a systematic analysis of Whisper's behavior in additive noise conditions and find that overconfident errors increase dramatically at low signal-to-noise ratios, with 10–20\% of tokens incorrectly predicted with confidence above 0.7. To mitigate this, we propose a lightweight, post-hoc calibration framework that detects potential overconfidence and applies temperature scaling selectively to those tokens, without altering the underlying ASR model. Evaluations on the R-SPIN dataset demonstrate that, in the low signal-to-noise ratio range ($-18$ to $-5$ dB), our method reduces the expected calibration error (ECE) by 58\% and triples the normalized cross entropy (NCE), yielding more reliable confidence estimates under severe noise conditions.

\end{abstract}


\section{Introduction}

Automatic speech recognition (ASR) systems now achieve high accuracy and are widely deployed, with data augmentation and robust training improving performance in noise. Yet beyond transcription accuracy, the reliability of model confidence---especially in end-to-end ASR---remains underexplored. Here, \textit{confidence} denotes the model’s estimated probability that a prediction is correct, typically the maximum posterior \cite{jiang2005confidence}; a well-calibrated model assigning 80\% confidence should be correct about 80\% of the time. Such reliability is critical for downstream tasks like error detection, selective re-recognition, and sample selection in semi-supervised learning (i.e., active learning). It is also essential in high-stakes domains such as legal or medical transcription.

Despite its importance, confidence calibration in end-to-end ASR under noisy acoustic conditions remains largely unexplored, due to several challenges. First, unlike traditional ASR systems with separate acoustic and language models, where post-hoc calibration is more accessible, end-to-end models rely on softmax scores that lack explicit confidence modeling. 
Second, while some work has examined ASR robustness patterns under accents or environmental shifts~\cite{graham2024evaluating, patman2024speech}, confidence patterns under noise remain unstudied, making it unclear how calibration should be adapted. Perceptual research on human listeners~\cite{huo2024release} shows that recognition patterns shift with acoustic cue degradation at different noise levels, underscoring the need for similar systematic analysis in ASR.
Finally, under noise, some errors are low-confidence and easy to reject, while others appear deceptively confident, raising the question whether calibration methods for clean speech generalize to noisy conditions, highlighting the need for focused study.

To address these challenges, we begin by analyzing confidence calibration in ASR under varying levels of additive noise. By injecting temporally modulated noise at varying signal-to-noise ratios (SNRs), we find that large pretrained models such as Whisper~\cite{radford2023robust} maintain reasonably good calibration on clean and moderately noisy inputs, even without explicit calibration during training. However, as noise intensifies, overconfidence becomes increasingly common, with models assigning high confidence to incorrect predictions. Based on these observations, we propose a post-hoc calibration framework that identifies when and where confidence estimates should be improved under noisy conditions.

This paper is organized as follows. Section~\ref{sec:related} reviews related work on confidence calibration for end-to-end ASR and general calibration techniques. Section~\ref{sec:performance} analyzes Whisper model's confidence reliability under noise. Section~\ref{sec:method} presents our post-hoc framework for selectively calibrating overconfident predictions. Section~\ref{sec:results} reports experimental results and ablations, and Section~\ref{sec:discussion} concludes with future directions.

\section{Related Works}
\label{sec:related}

\subsection{Robust ASR in Noisy Conditions}

Robustness to noise has long been a challenge in ASR. Early systems use speech enhancement front-ends~\cite{macho2002evaluation, yu2008minimum} but often suffer from mismatches between enhancement outputs and the recognition model’s acoustic assumptions. Later approaches integrate robustness directly into acoustic modeling, such as noise-aware training~\cite{seltzer2013investigation}. End-to-end ASR models now rely heavily on data augmentation strategies like multi-style training~\cite{ko2015audio}, SpecAugment~\cite{park2019specaugment}, and large-scale pretraining~\cite{baevski2020wav2vec, radford2023robust}. Multi-task training with auxiliary objectives (e.g., speech enhancement~\cite{chen2015speech}) further boosts robustness. 

While these methods improve recognition accuracy under noisy conditions, they do not explicitly address the reliability of confidence estimates.

\subsection{Confidence Calibration for ASR}

In traditional HMM-based ASR systems, confidence estimation starts with computing scores from scratch using decoder-internal information; the idea of calibrating these scores comes later. Such information includes word posteriors, alignments, and scores extracted from lattices or confusion networks~\cite{siu1997improved, evermann2000posterior,wessel2002confidence,cox2002high}. These methods are effective but remain tightly coupled to the decoder structure, limiting portability and system independence. To reduce dependency on internal components, later approaches leverage more accessible information, such as generic confidence measure from acoustic model and language model, word distributions, and rule coverage ratios~\cite{seigel2011combining, yu2011calibration}. While more portable, these ``semi-black-box" methods still assume explicit acoustic-language ASR architectures with interpretable intermediate features.

Modern end-to-end ASR systems, such as Whisper, eliminate explicit acoustic-language model boundaries and no longer generate lattices or other structured decoding outputs. As a result, traditional confidence calibration techniques are not applicable. Confidence is instead inferred from decoder softmax probabilities, marking a shift toward truly black-box estimation. However, softmax scores have the tendency to be overconfident without proper calibration~\cite{guo2017calibration}.

To improve confidence reliability in this setting, recent work has introduced confidence calibration modules (CEMs) that predict token- or word-level confidence from decoder logits, hidden representations and more~\cite{swarup2019improving,woodward2020confidence,li2021confidence, qiu2021learning, ogawa2021blstm}. These modules are typically trained on clean speech and have been shown to improve calibration performance. Extensions include multi-task CEMs that jointly model confidence, deletion prediction, and utterance-level quality~\cite{qiu2021multi}, fine-tuning Whisper decoder outputs for direct confidence prediction~\cite{aggarwal2025adopting}, and domain-adaptive approaches using pseudo-labels or external language model features~\cite{li2022improving}. A notable alternative is the entropy-based method~\cite{laptev2023fast}, which requires no training and estimates word-level confidence by aggregating uncertainty. Their method is among the few to demonstrate robustness under moderate noise conditions (0–30 dB SNR). While most efforts focus on word- or token-level estimation, some studies also explore utterance-level confidence scoring\cite{liu2021utterance, lee2024proper}.

\subsection{Temperature Scaling}

While the above studies focus on confidence calibration within ASR, related work in other domains offers broadly applicable techniques. One prominent example is temperature scaling, a post-hoc method that rescales logits by a single scalar to improve calibration without altering predicted classes, and is widely used in classification tasks such as image recognition and natural language processing~\cite{guo2017calibration, desai2020calibration}.

In ASR, temperature scaling has been explored only to a limited extent, either as a feature transformation~\cite{oneactua2021evaluation} or as a direct confidence correction method~\cite{woodward2020confidence}. These approaches target clean speech and overlook noisy conditions, where miscalibration is often worse. Our analysis in Section~\ref{sec:performance} shows that, under certain noise levels and contexts, miscalibrated tokens follow consistent patterns, suggesting the potential for targeted calibration. Similar ideas have been realized in other domains, where selective strategies apply temperature scaling only when it is most beneficial: Fisch et al.\cite{fisch2022calibrated} abstain from uncertain predictions, while Zollo et al.\cite{zollo2024improving} jointly train a calibrator and selector to adjust only easily correctable samples, achieving improved calibration in medical diagnosis.

Inspired by these works, we propose a selective calibration framework for end-to-end ASR that combines temperature scaling with a classifier to decide when and where calibration is most effective under noisy conditions.

\section{Confidence Behavior of ASR under Noise}
\label{sec:performance}

\begin{figure*}[t]
\centerline{\includegraphics[width=1.0\linewidth]{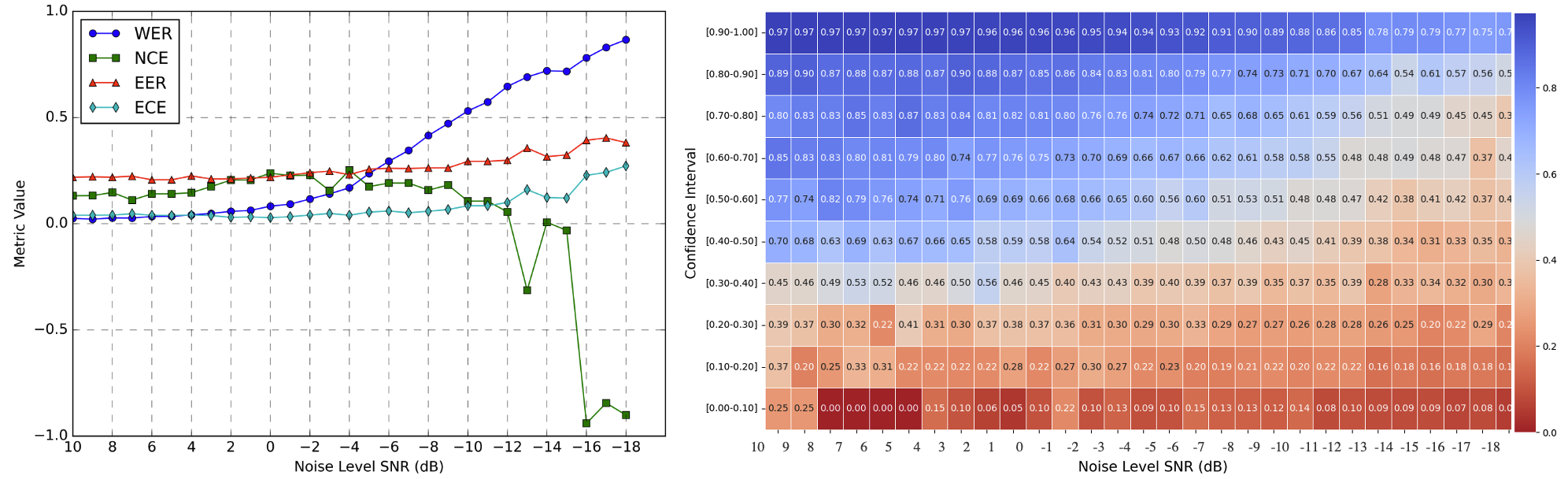}}
\caption{
Impact of additive noise level on ASR recognition and confidence performance. \textbf{Left}: recognition accuracy (WER ↓), calibration metrics (ECE ↓, NCE ↑), and discrimination metric (EER ↓). \textbf{Right}: correctness ratio per confidence bin across noise levels, where the percentage of correct tokens is presented both numerically and via color intensity.}
\label{fig:heatmap}
\end{figure*}

We evaluate the \texttt{Whisper-medium-en} model \footnote{https://huggingface.co/openai/whisper-medium.en}, a representative end-to-end ASR that fits our computational constraints. All audio is sampled at 16kHz.

\vspace{-3pt}
\subsection{Noise and Dataset Construction}
To simulate realistic acoustic degradation, we construct temporally modulated noise in two steps. First, white noise is filtered to match the long-term average spectrum of 500 utterances from the corpus, producing speech-shaped noise. Then, its amplitude is modulated using the average envelope of another 500 utterances to introduce natural temporal dynamics. The resultant noise signal retains the temporal modulation of actual speech without carrying any linguistic content, providing a controlled yet speech-like approximation of conversational backgrounds that enables consistent SNR settings while avoiding the unpredictability of real-world recordings.

For training, the temporally-modulated noise is then added to clean audio at 29 SNR levels from $+10$~dB to $–18$~dB (in 1~dB steps), with random onset and end to minimize alignment bias. Training uses LibriTTS clean-100~\cite{zen2019libritts} with diverse speakers and utterances, filtering out utterances shorter than 3~seconds or longer than 10~seconds. Each utterance is mixed with noise at one randomly selected SNR, without duplication across conditions.

Evaluation is performed on the R-SPIN corpus~\cite{wilson2012revised}, a well-established benchmark in phonetic and psychoacoustic research containing 400 phonetically balanced sentences. Although relatively small, it offers strong consistency and interpretability, with each sentence evaluated across all 29 SNR levels for controlled, sentence-aligned comparisons.


\subsection{Token-Level Confidence Extraction}
\label{sec:sec:conf_extraction}

Each noisy utterance is first transcribed using the Whisper model, and its word error rate (WER) is computed at the word level using the \texttt{jiwer} toolkit~\cite{morris2004and}. To ensure meaningful confidence analysis, we exclude obvious hallucination outputs using simple heuristics, including low average log-probability, high no-speech probability, or low compression ratio. 


Confidence analysis is performed at the token level, consistent with the model’s autoregressive decoding process. At each decoding step, we record the full logits vector (vocabulary size is 51,865) and compute the softmax confidence of the predicted token. A binary correctness label $y_i$ is assigned to the predicted token by comparing the prediction with the corresponding ground-truth token.

To characterize miscalibrated predictions, we define a token as \textit{overconfident} if it is incorrect ($y_i = 0$) but receives a softmax confidence of at least 0.7. This threshold is empirically chosen and remains consistent across SNR conditions. A binary overconfidence indicator is denoted as $o_i$.

\subsection{Metrics}

Besides recognition accuracy measured by WER, we evaluate ASR performance across several calibration metrics: expected calibration error (ECE), normalized cross entropy (NCE), negative log likelihood (NLL), and equal error rate (EER).

\begin{itemize}
    \item \textbf{WER} is computed at the word level across all utterances in an SNR level after removing hallucinated predictions, hence remaining below 1.
    \item \textbf{ECE}  divides predictions into $M=10$ equal-width confidence bins. For each bin \( B_m \), it measures the absolute difference between average confidence and accuracy:
        \begin{equation}
        \label{eq:ece}
        \text{ECE} = \sum_{m=1}^{M} \frac{|B_m|}{N} \left| \text{acc}(B_m) - \text{conf}(B_m) \right|,
        \end{equation}
        where \( N \) is the total number of tokens. ECE is between 0 and 1, and lower means better calibration.
    \item \textbf{NCE} captures the information gain from confidence scores compared to a non-informative baseline (i.e., dataset entropy). Following \cite{yu2011calibration}, we compute:
        \begin{equation}
        \label{eq:nce}
        \text{NCE} = \frac{H_{\text{base}} - H_{\text{cond}}}{H_{\text{base}}}
        \end{equation}
        \begin{equation}
            H_{\text{cond}} =  - \sum_{i=1}^{N} [y_i\log c_i + (1-y_i)\log(1-c_i)]
        \end{equation}
         
        \begin{equation}
            H_{\text{base}} = -n \log(\frac{n}{N}) - (N-n)\log(1-\frac{n}{N}).
        \end{equation}
    where \( c_i \) is the predicted confidence for token \( i \), \( y_i \) is the binary correctness label, $n$ and $N$ are numbers of correct and total tokens, respectively. Higher NCE indicates better calibration, while extremely poor calibration can result in negative values. While ECE reflects overall calibration, NCE penalizes overconfident outliers more severely.

    \item \textbf{NLL} is the unnormalized form of NCE and is commonly reported in calibration studies for comparability (lower is better).

    \item \textbf{EER} is the operating point where the false accept rate equals the false reject rate. Lower EER indicates better separability between correct and incorrect predictions, which is useful for decision-making scenarios.


    


\end{itemize}


\subsection{Observations}
As shown in Figure~\ref{fig:heatmap} left plot, all metrics degrade as noise intensifies (i.e., as SNR decreases from left to right), reflecting the overall decline in ASR performance. WER starts to increase rapidly below 0 dB, indicating a noticeable drop in recognition accuracy. In contrast, calibration metrics, ECE and NCE, remain relatively stable until around –5 to –10 dB, after which they degrade more sharply. This difference in turning points suggests that even when accuracy is suboptimal, the model still retains a degree of self-awareness (i.e., it knows when it is likely to be wrong), but this calibration breaks down under more severe noise conditions. 

The EER starts around 25\% even at high SNR levels, indicating that Whisper’s softmax probabilities are not strongly discriminative compared to traditional ASR confidence estimation modules. This is expected, as the model is not explicitly trained for it. Nevertheless, EER worsens with noise, reflecting reduced discriminative power in noisy conditions.


\begin{figure*}[t]
\centerline{\includegraphics[width=0.9\linewidth]{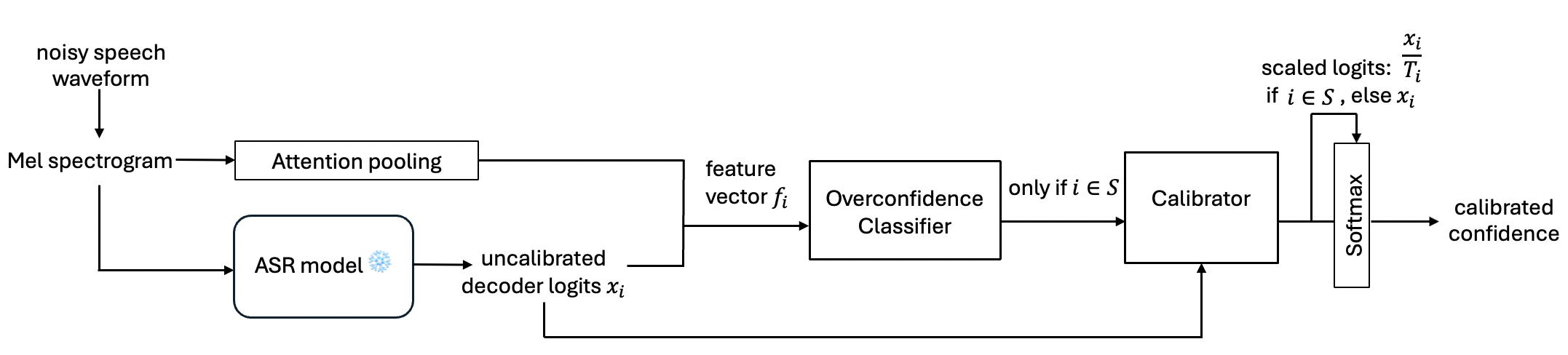}}
\caption{Overview of the proposed token-level selective calibration framework. A shared feature vector is computed for each token and passed through an overconfidence classifier. Only tokens predicted as overconfident are sent to the calibrator for temperature scaling, while others remain unchanged.}
\label{fig:model}
\end{figure*}


Notably, ECE captures overall miscalibration but cannot distinguish overconfidence from underconfidence due to its use of absolute value in \eqref{eq:ece}. To explicitly examine the direction of miscalibration, we analyze token accuracy across confidence bins and SNR levels. The right plot in Figure~\ref{fig:heatmap} provides a fine-grained view, showing that under severe noise, high-confidence tokens (e.g., 0.7–1.0) become increasingly incorrect, highlighting overconfidence that ECE alone cannot reveal.

\section{Proposed Method}
\label{sec:method}

\subsection{Overview}
\label{sec:method-overview}
Motivated by the prevalence of overconfidence under noise, we propose a two-stage framework to improve ASR reliability by selectively calibrating such predictions. It consists of (1) an overconfidence classifier and (2) a temperature-based calibrator. The structure is shown in Figure~\ref{fig:model}.

Both modules operate on a shared feature vector $\mathbf{f}_i$ constructed for each token \(i\). These features can be grouped into three categories heuristically, reflecting different aspects of model behavior:
\begin{itemize}
    \item \textbf{Confidence uncertainty}: top-1 probability, top-2 margin, and entropy from the softmax distribution.
    \item \textbf{Token plausibility}: predicted token ID embedding, positional index, and top-$k$ raw logits.
    \item \textbf{Acoustic clarity}: a global utterance-level embedding derived from the Mel spectrogram via attention pooling, and shared by all tokens in the utterance.
\end{itemize}


\subsection{Overconfidence Classifier}

We use a binary classifier to detect overconfident predictions based on per-token feature vectors $f_i$. The classifier outputs a probability $\hat{o_i} \in [0, 1]$, indicating the likelihood that token $i$ is overconfident. Ground-truth labels $o_i\in\{0,1\}$ are derived by comparing the token’s softmax confidence with a threshold and its recognition correctness $y_i$, as described in Section~\ref{sec:sec:conf_extraction}. 

To address the class imbalance between overconfident and non-overconfident tokens, we apply a weighted binary cross-entropy loss $\mathcal{L}_\text{BCE}$	with a positive class weight of $w=7$, selected empirically. The classifier is implemented as a simple two-layer multilayer perceptron (MLP) and produces sufficiently informative outputs to guide the downstream calibrator. 

\begin{equation}
\mathcal{L}_{\text{BCE}} = - \frac{1}{N} \sum_{i=1}^{N} \left[ w \cdot o_i \log \hat{o}_i + (1 - o_i) \log (1 - \hat{o}_i) \right]
\end{equation}

\begin{table*}[t]
\centering
\caption{
Calibration results on the low-SNR range ($-$18 to $-$5 dB) of the R-SPIN dataset. All metrics are computed over all tokens ($n = 25{,}450$). “ECE Improvement” indicates reduction from the uncalibrated baseline. Overconfident mass is the proportion of incorrect tokens assigned confidence above 0.7.
}
\label{tab:main}
\begin{tabular}{|l|c|c|c|c|c|c|}
\hline
\textbf{Method} & \textbf{ECE} $\downarrow$ & \textbf{ECE Improv.} $\uparrow$ & \textbf{NLL} $\downarrow$ & \textbf{NCE} $\uparrow$ & \textbf{EER} $\downarrow$ & \textbf{Overconf. Mass} $\downarrow$ \\
\hline
No calibration                 & 0.086 & –      & 5.910 & 0.064  & 28.8\% & 11.1\% \\
SNR-based calibration           & 0.082 & 0.004  & 5.688 & 0.082 & \textbf{27.4\%} & 9.0\%  \\
\textbf{Selective calibration (ours, token-level T)} & \textbf{0.036} & \textbf{0.050} & \textbf{5.651} & \textbf{0.192} & 27.8\% & 6.6\%  \\
\hspace{2em}Utterance-level T & 0.040 & 0.046 & 5.651 & 0.191 & 27.9\% & \textbf{6.5\%} \\
\hline
\end{tabular}
\end{table*}

\subsection{Selective Calibrator}

To recalibrate confidence selectively, we apply temperature scaling at the token level, but only to tokens predicted as overconfident from the previous stage. This targeted approach focuses on correction where needed while preserving well-calibrated predictions elsewhere.

Let \(\mathcal{S} = \{ i \mid \hat{o}_i \ge 0.5 \}\) denote the set of tokens flagged as overconfident by the classifier. For each \(i \in \mathcal{S}\), let \(\mathbf{x}_i\) be the uncalibrated logits, \(y_i\) the ground-truth correctness label, and \(T_i > 1\) the predicted temperature. Scaling is applied as:

\begin{equation}
\mathbf{x}_i^{\text{scaled}} = \frac{\mathbf{x}_i}{T_i}
\end{equation}

The temperature \(T_i\) is predicted from the same feature vector \(\mathbf{f}_i\) using a small feedforward network, where $W_2$, $W_1$, $b_1$ and $b_2$ are learnable parameters:

\begin{equation}
T_i = 1 + \text{softplus}(W_2 \cdot \text{ReLU}(W_1 \mathbf{f}_i + \mathbf{b}_1) + b_2)
\end{equation}

The use of \texttt{softplus} and an additive offset ensures \(T_i > 1\), guaranteeing that scaling only reduces confidence and never inflates it. The full sequence of logits is then reassembled as:

\begin{equation}
\hat{\mathbf{x}}_i = 
\begin{cases}
\mathbf{x}_i^{\text{scaled}}, & \text{if } i \in \mathcal{S} \\
\mathbf{x}_i, & \text{otherwise}
\end{cases}
\end{equation}
The calibrator module is optimized with two loss terms. First, a cross-entropy (CE) loss is applied only to the recalibrated tokens in \(\mathcal{S}\), encouraging more accurate predictions after calibration:

\begin{equation}
\mathcal{L}_{\text{CE}} = \frac{1}{|\mathcal{S}|} \sum_{i \in \mathcal{S}} \text{CE}\left(\textbf{x}_i^{\text{scaled}},\; y_i \right)
\end{equation}

Second, a \textit{soft-binned} expected calibration error (ECE) loss~\cite{karandikar2021soft}, which replaces \textit{hard binning} of confidence intervals with a continuous formulation via a trainable soft bin membership function, is computed over the full reassembled sequence to optimize global calibration:

\begin{equation}
\mathcal{L}_{\text{ECE}} = \text{ECE}_{\text{soft-binned}}(\{ \hat{\mathbf{x}}_i \}_{i=1}^{n}, \{ y_i \}_{i=1}^{n})
\end{equation}

\subsection{Joint Training}
To encourage the classifier to produce more calibration-relevant predictions, we train both modules jointly using the shared input feature \(\mathbf{f}_i\) and a combined loss:
\begin{equation}
    \mathcal{L}_{\text{total}} = \lambda_{\text{BCE}} \cdot \mathcal{L}_{\text{BCE}} + \lambda_{\text{CE}} \cdot \mathcal{L}_{\text{CE}} + \lambda_{\text{ECE}} \cdot \mathcal{L}_{\text{ECE}}
\end{equation}

We set the loss weights via grid search on the validation set: \(\lambda_{\text{BCE}}=0.5\), \(\lambda_{\text{CE}}=1.0\), and \(\lambda_{\text{ECE}}=10.0\). During training, the classifier flags 10–15\% of tokens per batch as overconfident, closely matching the ground-truth distribution in the dataset. Although its standalone performance is modest (recall~$\sim$0.7, precision~$\sim$0.3, F1~$\sim$0.4), high recall is especially valuable for this task. Notably, freezing the classifier degrades calibration performance, confirming the benefit of joint optimization. The full model remains lightweight, with only 3.4M parameters. To focus on challenging cases, we train using only LibriTTS utterances from the low SNR range ($-18$ to $-5$~dB).



\section{Results}
\label{sec:results}

\subsection{Improved Calibration Results}
Table~\ref{tab:main} reports calibration results on the low-SNR range ($-18$ to $-5$ dB) of the R-SPIN dataset. Without calibration, the Whisper model exhibits substantial overconfidence, with over 11\% of tokens being incorrect yet assigned high confidence scores. Our proposed token-level selective calibration achieves the largest improvement in expected calibration error (ECE), reducing it by 0.050 absolute (a 58\% relative reduction), and lowers the overconfident mass from 11.1\% to 6.6\% on average. Since the method is not explicitly designed to improve discriminative power, the EER remains largely unchanged.

We compare against a non-trainable, non-selective baseline that groups tokens by SNR and applies a grid-searched temperature per SNR level. While this method reduces ECE moderately, it underperforms our approach and requires explicit knowledge of the input SNR, limiting its applicability.

To further evaluate the effect of calibration granularity, we fix the overconfidence classifier and compare token-level calibration with a coarser variant that predicts one temperature per utterance. As shown in Table~\ref{tab:main} last two rows, token-level scaling consistently outperforms utterance-level across calibration metrics, benefiting from finer-grained control under noisy conditions. Nonetheless, utterance-level calibration still offers reasonable gains with reduced complexity.

\subsection{SNR-specific Performance}
Table~\ref{tab:snr} breaks down calibration performance across different SNR ranges. Our token-level selective calibration consistently improves ECE, NCE, and overconfidence mass in challenging conditions (SNR $\le –5$ dB). Importantly, the method also preserves calibration in high-SNR settings (SNR $> 5$ dB), avoiding unnecessary correction when predictions are already reliable. In mid-SNR ranges (i.e., $–5$ to $+5$ dB), our method brings limited improvements and may introduce mild degradation---yet overall reliability remains stable.

\begin{figure}[htbp]
\centerline{\includegraphics[width=0.99\linewidth]{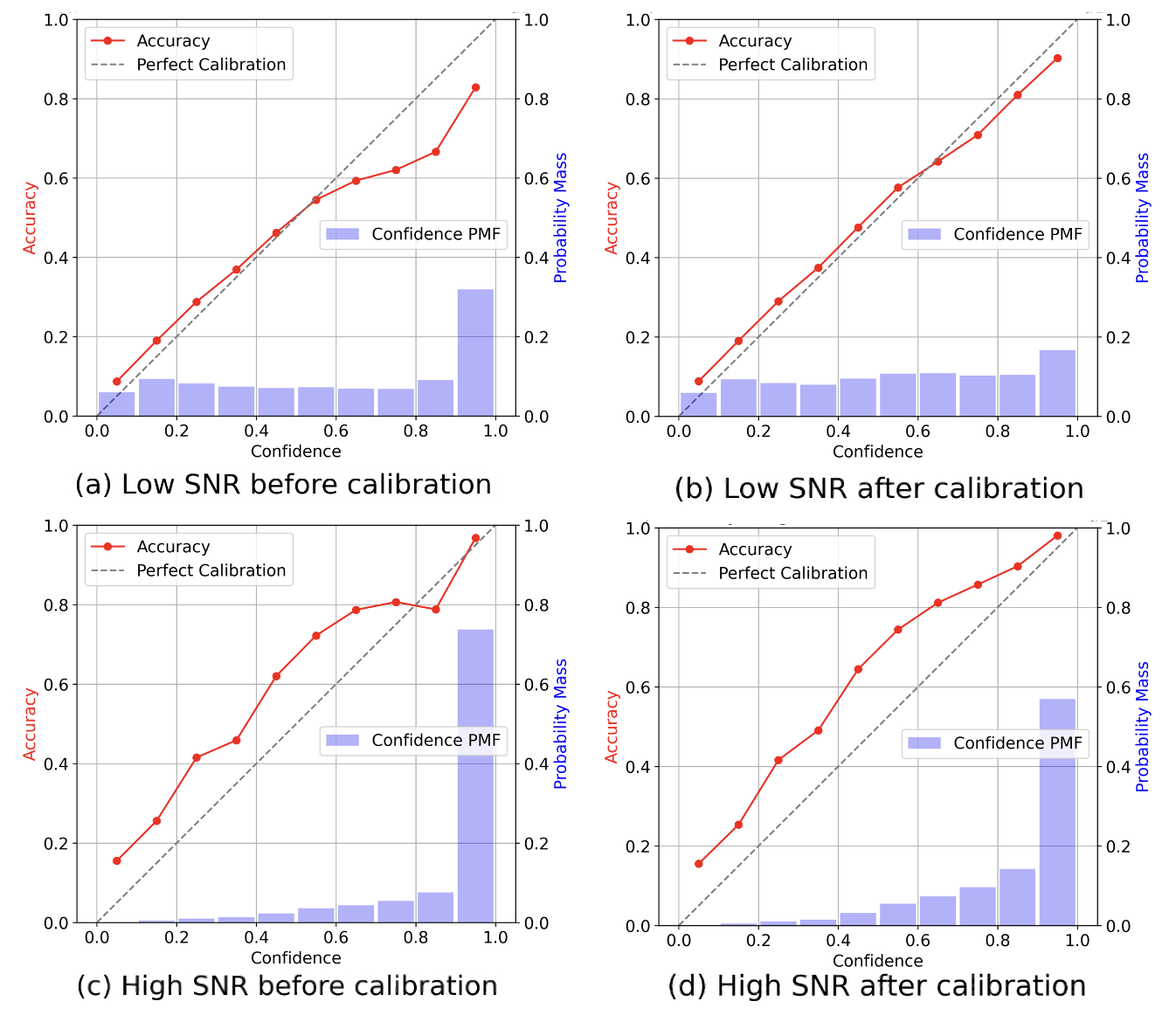}}
\caption{Reliability diagrams before and after selective calibration under low-SNR ($-18$ to $-5$ dB) and high-SNR ($-4$ to $+10$ dB) conditions. Red lines plot accuracy per confidence bin; the dashed diagonal indicates perfect calibration. Blue bars show the confidence score distribution.}

\label{fig:reliability}
\end{figure}

Figure~\ref{fig:reliability} shows reliability diagrams for low- and high-SNR regimes. At low SNR, our method reduces overconfidence and aligns predicted confidence with empirical accuracy (grey diagonal line). At high SNR, calibration remains largely intact, with only a slight shift from high- to mid-confidence bins, indicating that the method avoids over-correction when predictions are already reliable.

\begin{table}[t]
\centering
\caption{Calibration results at SNR ranges. ``Before" denotes the uncalibrated baseline, ``After" shows the effect of our token-level selective calibration method.}
\label{tab:snr}

\begin{tabular}{|c|l|c|c|c|}
\hline
\textbf{SNR (dB)} & \textbf{Method} & \textbf{ECE} $\downarrow$ & \textbf{NCE} $\uparrow$ & \textbf{Overconf. Mass} $\downarrow$ \\
\hline
\multirow{2}{*}{(5,10]} 
    & Before         & 0.0412 & 0.1329  & 5.0\% \\
    & After          & \textbf{0.0335} & \textbf{0.1537}  & \textbf{3.9\%} \\
\hline
\multirow{2}{*}{(0,5]} 
    & Before         & \textbf{0.0361} & \textbf{0.1773}  & 5.0\% \\
    & After          & 0.0460 & 0.1388  & \textbf{3.9\%} \\
\hline
\multirow{2}{*}{(-5,0]} 
    & Before         & \textbf{0.0381} & \textbf{0.2239}  & 5.1\% \\
    & After          & 0.0755 & 0.1589  & \textbf{3.8\%} \\
\hline
\multirow{2}{*}{[-10,-5]} 
    & Before         & 0.0573 & 0.1728  & 8.1\% \\
    & After          & \textbf{0.0517} & \textbf{0.1988}  & \textbf{5.4\%} \\
\hline
\multirow{2}{*}{[-15,-10)} 
    & Before         & 0.1115 & -0.0235  & 12.7\% \\
    & After          & \textbf{0.0611} & \textbf{0.1117}  & \textbf{7.6\%} \\
\hline
\multirow{2}{*}{[-18,-15]} 
    & Before         & 0.2396 & -0.8933 & 22.4\% \\
    & After          & \textbf{0.1349} & \textbf{-0.0511} & \textbf{10.2\%} \\
\hline
\end{tabular}
\label{tab:snr_metrics}
\end{table}


\subsection{Ablation Study on Features}

\textbf{Acoustic representations.}  
We compare alternative acoustic inputs for the modules. Replacing Mel spectrograms with frozen Whisper encoder states yields a slightly better ECE (0.0502 → 0.0507), confirming the usefulness of high-level features. However, this increases training time and memory, so we retain Mel features with attention pooling in the final model. Surprisingly, using only the first 3 seconds of audio consistently outperforms the full-length input, likely because early frames already capture the acoustic distortion. We adopt this truncated setting by default.

\textbf{Leave-one-out analysis.}  
To assess the contribution of each feature, we perform leave-one-out ablation on the final feature set. As shown in Table~\ref{tab:ablation}, most removals degrade calibration. Acoustic and semantic cues (Mel, token embedding) are particularly important, and removing uncertainty-based signals (entropy, margin) also hurts performance, confirming their complementarity in detecting overconfident errors~\cite{laptev2023fast}.

Among all features, the top-$k$ logits ($k=5$) stand out. They capture uncertainty among multiple plausible predictions, beyond the top-1 confidence. Removing this feature superficially improves ECE on the validation set, but further analysis on the phonetically balanced evaluation set reveals \textit{overcorrection}: the model begins adjusting already well-calibrated predictions, reducing generalization under high SNR. This highlights the value of fine-grained uncertainty signals like top-$k$ logits for targeted and reliable calibration.

\begin{table}[t]
\centering
\caption{Feature ablation results on ECE improvement in the low-SNR range ($-$18 to $-$5 dB), comparing to the uncalibrated baseline.}
\label{tab:ablation}
\begin{tabular}{|l|c|}
\hline
\textbf{Feature Variant} & \textbf{ECE Improv. $\uparrow$} \\
\hline
Full audio Mel + avg pooling & 0.0481 \\
Full audio Mel + attention pooling & 0.0493 \\
\textbf{3s Mel + attention pooling (final model)} & 0.0502 \\
Whisper encoder hidden states & \textbf{0.0507} \\
\hline
No top-1 confidence & 0.0495 \\
No margin & 0.0502 \\
No entropy & 0.0503 \\
No position & 0.0494 \\
No top-$k$ logits & 0.0476 \\
No mel spectrogram & 0.0485 \\
No token embedding & 0.0441 \\
No mel + token emb & 0.0457 \\
No entropy + margin & 0.0477 \\
\hline
\end{tabular}
\end{table}

\vspace{-3pt}

\section{Discussion}
\label{sec:discussion}

\subsection{Auxiliary Features: SNR}
Given the strong influence of noise level on calibration performance, we explored whether incorporating explicit acoustic quality (utterance-level SNR) could further improve results. In an oracle setting, ground-truth SNR values were appended to the model input using either direct concatenation or gated modulation. These were applied to the classifier, the calibrator, or both. While slight improvements in ECE were observed, the gains were modest and inconsistent across conditions.

We also evaluated a predicted-SNR variant, using a MLP trained jointly via weighted mean squared error. Interestingly, performance was comparable to the oracle setting, suggesting that SNR prediction may serve as a task-adaptive latent signal. However, none of the SNR-based variants consistently outperformed the existing model. These results suggest that utterance-level acoustic quality is likely already encoded in the Mel spectrogram features. As explicit SNR injection offered limited value, we exclude it from the final model.






\subsection{Token-Level Acoustic Alignment}

While most features are token-specific, the acoustic embedding derived from Mel spectrograms is utterance-level, obtained via attention pooling over the first 3 seconds. Despite this mismatch in granularity, ablation results indicate that this coarse acoustic summary is sufficient for effective calibration, even outperforming full-length representations.

We hypothesize two contributing factors. First, the acoustic embedding may primarily capture global properties such as background noise level or modulation patterns, which influence calibration but do not require fine-grained temporal precision. In contrast, token-level features directly reflect decoder uncertainty and dominate the calibration information.

Second, the use of synthetic, temporally-modulated noise creates relatively uniform acoustic conditions within each utterance, reducing the need for local alignment. In more dynamic real-world scenarios, e.g., streaming audio with abrupt noise changes, finer-grained acoustic modeling may be more effective. Future work could explore token-aligned features or sliding-window embeddings to improve robustness.



\section{Conclusion}

We studied the calibration behavior of end-to-end ASR models under noise, using \texttt{Whisper-medium-en} as a case study. Although well-calibrated on clean inputs, Whisper becomes increasingly overconfident under noise, assigning high confidence to incorrect predictions. To address this, we proposed a lightweight post-hoc framework that identifies overconfident tokens using acoustic, uncertainty, and contextual features, and selectively applies temperature scaling. This improves calibration without modifying the ASR model itself. Evaluations on the R-SPIN dataset show that our method reduces ECE and improves NCE in low-SNR settings, while preserving calibration in high-SNR cases. The calibrator is modular and token-adaptive; while we use temperature scaling here, the method can extend to other differentiable strategies.  Future work may explore finer acoustic alignment to handle real-world, dynamic noise conditions more effectively.

\section*{Acknowledgment}
This work used the Delta system at the National Center for Supercomputing Applications through allocation CIS250338 from the Advanced Cyberinfrastructure Coordination Ecosystem: Services \& Support (ACCESS) program \cite{boerner2023access}, which is supported by U.S. National Science Foundation grants \#2138259, \#2138286, \#2138307, \#2137603, and \#2138296.

\bibliographystyle{IEEEtran}
\bibliography{references}


\end{document}